\documentclass[aps,prb,twocolumn,preprintnumbers,amsmath,amssymb,superscriptaddress]{revtex4}%

\usepackage{graphicx}%
\usepackage{dcolumn}
\usepackage{amsmath}
\usepackage{color}
\usepackage{ulem}

\newcommand{\RN}[1]{\textup{\uppercase\expandafter{\romannumeral#1}}}%

\begin{document}

\title{Optical setup for a piston-cylinder type pressure cell: a double-volume approach}

\author{Pavel Naumov}
 \affiliation{Quantum Criticality and Dynamics Group, Paul Scherrer Institut, CH-5232 Villigen PSI, Switzerland}

\author{Ritu Gupta}
 \affiliation{Laboratory for Muon Spin Spectroscopy, Paul Scherrer Institute, CH-5232 Villigen PSI, Switzerland}

\author{Marek Bartkowiak}
 \affiliation{Laboratory for Neutron and Muon Instrumentation, Paul Scherrer Institute, CH-5232 Villigen PSI, Switzerland}

\author{Ekaterina Pomjakushina}
 \affiliation{Laboratory for Multiscale Materials Experiments, Paul Scherrer Institut, CH-5232 Villigen PSI, Switzerland}

\author{Nicola P. M. Casati}
 \affiliation{Laboratory for Synchrotron Radiation – Condensed Matter, Paul Scherrer Institut, CH-5232 Villigen PSI, Switzerland}

\author{Matthias Elender}
 \email{matthias.elender@psi.ch}
 \affiliation{Laboratory for Muon Spin Spectroscopy, Paul Scherrer Institute, CH-5232 Villigen PSI, Switzerland}

\author{Rustem Khasanov}
 \email{rustem.khasanov@psi.ch}
 \affiliation{Laboratory for Muon Spin Spectroscopy, Paul Scherrer Institute, CH-5232 Villigen PSI, Switzerland}

\begin{abstract}
Measurement of the absolute value of the applied pressure in high-pressure muon and neutron experiments is a complicated task. It requires both the presence of a calibration material inside the sample volume as well as additional time for refining the response of the calibrant. Here we describe the use of optical calibrants for precise determination of the pressure value inside the piston-cylinder clamp cells.  Utilizing the concept of separate volumes for the sample and the optical media, a new setup for conducting in-situ pressure measurements has been successfully tested. Pressures in both the `sample' and the `optical' volumes were proved to be the same within experimental accuracy. The use of SrB$_{4}$O$_7$:(0.01 Sm$^{2+}$, 0.03 Eu$^{2+}$) as a pressure calibrant allows for a high accuracy of pressure determination by considering up to eight fluorescence lines.

\end{abstract}

\maketitle

\section{Introduction}\label{sec:introduction}

Experiments under high-pressure represent an integral field in condensed matter physics.  Currently,  large neutron and muon research user facilities are equipped with various types of high-pressure devices allowing for experiments in high magnetic fields (up to several tens of Tesla) and low temperatures (down to few millikelvin). Among the many different types of cells, one of the most popular remains the clamp-type construction, which allows one to reach pressures as high as $\simeq 3$~GPa and is characterised by a reasonably large volume of the sample chamber compared to its compact pressure cell dimensions.\cite{Eremets_book_1996, Klotz_book_2013}

Measurement of the absolute value of the pressure inside the piston-cylinder clamp cells used in neutron and muon-spin rotation/relaxation ($\mu$SR) experiments is a complicated task. Generally, the sample is loaded, pressurized and locked at room temperature outside of the neutron/muon beam. Pressure determination in clamp cells are separated, therefore, in `ex-situ' and `in-situ' measurements.

For `ex-situ' measurements, the pressure is determined outside of the neutron/muon spectrometer. There are two possible ways to make the pressure determination: (i) the 'contact' or (ii) the `contact-less' methods are generally considered. In Case (i), the connection (contact) wires enter the cell.\cite{Eremets_book_1996, Klotz_book_2013} The pressure is typically determined by measuring resistivity of a manganin wire, or the superconducting transition of some elemental metals such as Sn, In, or Pb.\cite{Eremets_book_1996, Klotz_book_2013, Wang_RSI_2011, Eiling_IJMP_1981} The disadvantage of the `contact'-type of pressure determination is that one needs to introduce the current leads inside the volume of the cell where the high-pressure is generated. Therefore, special attention must be given in the preparation of the feedthrough which must allow the current leads to enter but also must be strong enough to not be blown out by the pressure generated inside the cell.\cite{Eremets_book_1996, Walker_RSI_1999}
In case (ii), {\it i.e.} in the 'contact-less' case, the pressure is determined by measuring the AC response of elemental superconductors as Sn, In, or Pb,\cite{Khasanov_HPR_2016, Shermadini_HPR_2017} the optical response of the fluorescence material,\cite{Podlesnyak_HPR_2018} or the enhancement of the pressure cell outer diameter.\cite{Shermadini_HPR_2017} The disadvantage of a 'contact-less' case is that one needs either to find a way to access the pressure probe (optical or superconducting) via the clamped cell body or calibrate the enhancement of the cell diameter as a function of pressure inside the cell.

The `in-situ' pressure determination in neutron experiments is typically performed by measuring the unit-cell parameters of a material with a known equation of state (such as NaCl, MgO, or Pb) mixed with the sample.\cite{Klotz_book_2013, Decker_JAP_1971}  This method has several disadvantages, since the calibrant material reduces much needed sample volume, may absorb neutrons, or could even react with the sample. In $\mu$SR studies the `in-situ' pressure determination was possible only in two experiments, where the sample itself (Al in Ref.~\onlinecite{Khasanov_Al_PRB_2021} and Sr$_2$RuO$_4$ in Ref.~\onlinecite{Khasanov_SRO_unp}) was used as the pressure calibrant.

\begin{figure*}[htb]
\centering
\includegraphics[width=0.8\linewidth]{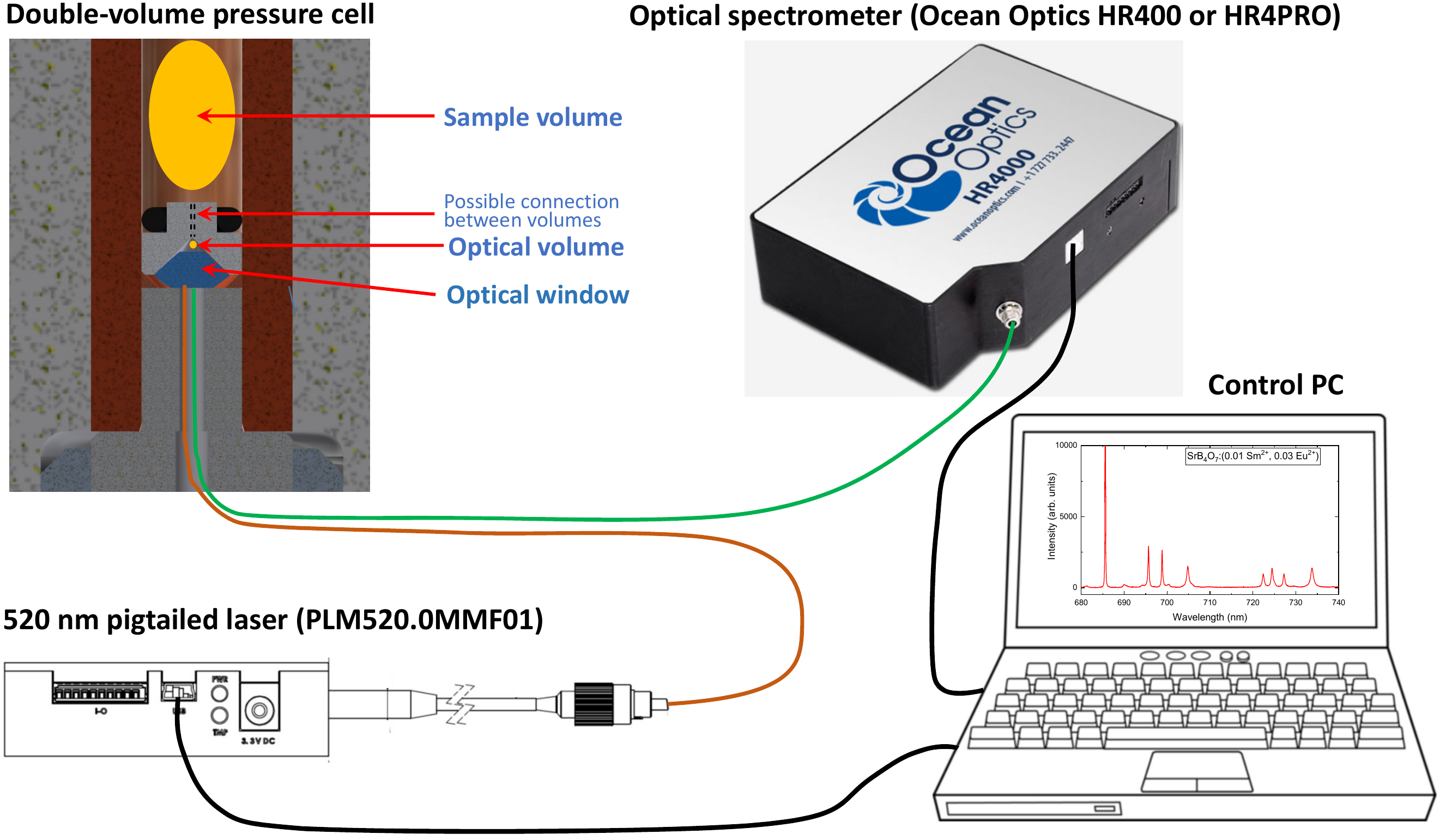}
\caption{The concept and the pressure measuring principle of the double-volume clamp cell. The pressure cell (top left corner) has two separate volumes: the first for the sample (large yellow oval) and the second for the optical probe (small yellow circle). The volumes may be connected via a channel. The measurement setup includes the laser source (PLM-520.0-MMF-0.1  Solid-State Laser from `Lasercomponents', Ref.~\onlinecite{Laser_lasercomponents}), the optical spectrometer (HR4000 or HR4PRO from Ocean Optics, Ref.~\onlinecite{Spectrometer_OceanOptics}) and the control PC (with the 'Ocean View`, Ref.~\onlinecite{Spectrometer_OceanOptics}, and the laser operation, Ref.~\onlinecite{Laser_lasercomponents}, software). The pressure cell, the excitation laser, and the measuring spectrometer are connected with each other via the optical reflection probe (Art Photonics, Ref.~\onlinecite{artphotonics}). }
 \label{fig:concept_of_double-volume pressure cell}
\end{figure*}

It is thus crucial to monitor the pressure inside the cell without reducing the volume available for the sample. It could also be important, for some experiments, to prevent a mixing of the sample and the pressure indicator materials. For resolving the above issues, the concept of a double-volume pressure cell, where the 'optical` and the 'sample` volumes are physically separated from each other, was developed and successfully tested. The construction of the piston-cylinder cell was modified by introducing the tungsten carbide piston with the entrance for the optical fibers, the teflon seal with the conical cavity, and the optical window made of commercially available cubic-Zirconia single crystals.

The paper is organized as follows: Section~\ref{sec:Motivation} describes the concept of a double-volume piston-cylinder clamp cell and the optical setup for pressure determination.  The description of the double-volume pressure cell assembly is given in Section~\ref{sec:Cell-Assembly}. The parts of the optical setup,  the optical window and the teflon mushroom, are discussed in Sections~\ref{sec:Optical-Window} and \ref{sec:Teflon-Musroom}, respectively. The results of calibration of the optical sensor [SrB$_{4}$O$_7$:(0.01 Sm$^{2+}$, 0.03 Eu$^{2+}$)] are presented in Section~\ref{sec:SrB4O7}. Section~\ref{sec:experiments} describes the test results of the double-volume pressure cell. The finite element analysis data are described in Section~\ref{sec:FEA}. Conclusions follow in Section~\ref{sec:conclusions}.

\section{The concept of the double-volume clamp cell and measuring principle}
\label{sec:Motivation}

In a double-volume clamp cell design, the inner pressure cell volume is divided into sample and the sensor volumes, respectively. The name 'double-volume` therefore describes the physical separation of the two volumes (the first occupied by the sample and the second by the pressure probe). An example of such design is presented at the top left corner of Fig.~\ref{fig:concept_of_double-volume pressure cell}, where the large yellow oval and the small yellow circle represent the sample and the pressure probe volume, respectively. The crucial condition of such design is that the partitioning material transmits the pressure perfectly.

The double-volume clamp cell has several advantages since: \\
(i) The calibrant material does not mix with the sample, so any possible reaction between the sample and the calibrant is not possible.\\
(ii) The optical probe stays away from the sample region and will thus not be hit by neutrons or muons with properly collimated beams. \\
(iii) The volume occupied by the optical probe is far smaller compared to the sample volume. Indeed, the fluorescence measurements could be easily performed on a small ruby crystal ($\sim0.01$~mm$^3$ or less), while the 'sample` volume for a typical piston-cylinder clamp cell used in neutron/muon experiments is of the order of 200-500~mm$^3$.\cite{Khasanov_HPR_2016, Shermadini_HPR_2017, Podlesnyak_HPR_2018, Sadykov_JPCS_2017, Sadykov_JNR_2018, Khasanov_ThreeWall_Arxiv_2021, Grinenko_NatCom_2021}\\
(iv) Both the `sample' and the `optical' volumes could be connected via a channel drilled inside the pressure seal (thin dashed lines in Fig.~\ref{fig:concept_of_double-volume pressure cell} marked as `possible connection between volumes').

The measurement setup, as is shown in Fig.~\ref{fig:concept_of_double-volume pressure cell}, includes the laser source, the optical spectrometer, the optical reflection probe, and the controlling PC. In our experiments, the excitation light was provided by a 520~nm PLM-520.0-MMF-0.1  Solid-State Laser from `Lasercomponents' with a maximum power of 25~mW.\cite{Laser_lasercomponents} The fluorescence spectra were measured using  HR4000 or HR4PRO spectrometers from Ocean Optics.\cite{Spectrometer_OceanOptics} The optical reflection probe with the 1.4~mm ferrule terminated fibers was obtained from the Art Photonics.\cite{artphotonics} The control PC has the 'Ocean View' and laser operation software installed.\cite{Spectrometer_OceanOptics, Laser_lasercomponents}

\section{The double-volume pressure cell assembly}\label{sec:Cell-Assembly}

\begin{figure}[htb]
\centering
\includegraphics[width=1.0\linewidth]{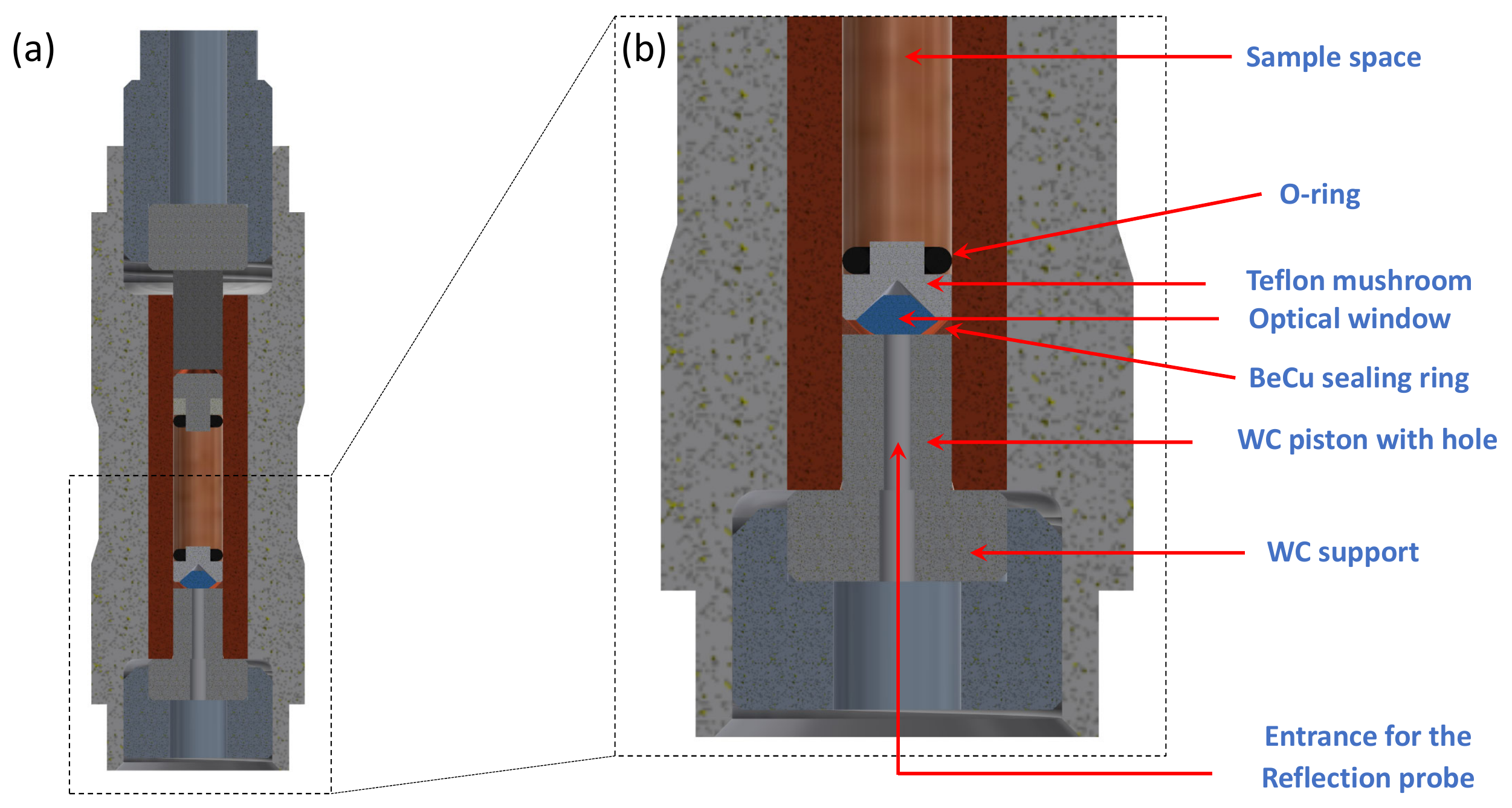}
\caption{(a) Cross-sectional view of the double-volume piston-cylinder pressure cell. The pressure cell body (with the inner and the outer diameters $\varnothing6$ and $\varnothing24$~mm, respectively) and the top part of the pressure seal [the mushroom, the tungsten carbide (WC) piston, the WC support, and the fixing bolt] are the same as described in Refs.~\onlinecite{Khasanov_HPR_2016, Shermadini_HPR_2017, Khasanov_ThreeWall_Arxiv_2021}. (b) The expanded part of the cell with the optical setup. The main elements are: the teflon mushroom with the conical entrance, the optical window prepared from the commercial gem crystal with the `brilliant-cut' shape, and the WC piston with the entrance for optical fibers. }
 \label{fig:double-volume pressure cell}
\end{figure}

A schematic view of a double-volume clamp cell is presented in Fig.~\ref{fig:double-volume pressure cell}~(a). The pressure cell body (with the inner and the outer diameters $\varnothing6$ and $\varnothing24$~mm, respectively) and the top part of the pressure seal [the mushroom, the tungsten carbide (WC) piston, the WC support pad, and the fixing bolt] are the same as currently used in clamped cells for muon-spin rotation/relaxation experiments,\cite{Khasanov_HPR_2016, Shermadini_HPR_2017, Khasanov_ThreeWall_Arxiv_2021} and are not discussed here.

The expanded view of the `optical' part of the double-volume cell is presented in Fig.~\ref{fig:double-volume pressure cell}~(b). It consists of the $\varnothing12$~mm WC support disk with a 2~mm hole, the  $\varnothing6$~mm WC piston with a 1.5~mm hole, the optical window made out of a 4~mm single crystalline piece of cubic-Zirconia, and the teflon mushroom with the conical entrance for the optical window. The optical probe ($\simeq 0.1-0.5$~mm$^3$ crystal of ruby and/or a small amount of  SrB$_{4}$O$_7$ powder), is placed on the top of the optical window together with a small drop of the pressure transmitting media (Daphne 7373 oil, in our case).

Here below we describe separately the main elements of the double-volume optical setup, namely the optical window and the teflon mushroom.

\subsection{Optical window}\label{sec:Optical-Window}

The optical window is used to transfer excitation light and collect the reflected light, which contains the fluorescence response of the optical probe (see Fig.~\ref{fig:concept_of_double-volume pressure cell}). This means that the window needs to both be transparent to light and also strong enough to withstand high pressure.

In the vast majority of pressure experiments, the media used for optical windows are sintered diamond.\cite{Eremets_book_1996, Klotz_book_2013, Podlesnyak_HPR_2018, Almax} Recently, cubic-Zirconia  (c-Zirconia) and Moissanite single-crystals have become easily available. They are machined in a so-called `brilliant-cut' shape and are often used in the jewelry industry as a low-cost substitute for diamonds.

\begin{table}[htb]
     \centering
     \caption{Comparison of mechanical properties of different gem stones after Refs.~\onlinecite{Xu_JPCM_2002,Dubrovinsky_NatCom_2012}. The prices are taken from Refs.~\onlinecite{Aliexpress, Almax}. The row `highest pressure' refers to the maximum pressure for the materials used as an anvil in anvil-type pressure cells. \\ }
     \begin{tabular}{c|cccc}
  & Diamond & Moissanite & Sapphire & c-Zirconia \\
 \hline
Composition  & C & SiC & Al$_{2}$O$_{3}$ & ZrO$_{2}$ \\
Hardness(Mohr)  & 10 & 9.25 &	9 & 8.5 \\
Highest $p$ (GPa) & 640 & 58.7	& 25.8 & 16.7 \\
Price (USD)  &800-1200&10 &20 &	0.2  \\
     \end{tabular}
     \label{tab:Anvils}
 \end{table}

Comparison of mechanical properties of different gem stones used in high-pressure experiments are presented in Table~\ref{tab:Anvils}. Here, the `Highest pressure' row refers to the maximum pressure each material can withstand when used as an anvil in anvil-type pressure cells.  Following the results presented in Table~\ref{tab:Anvils}, even cubic-Zirconia is suitable for use as an optical window in clamp-type cells, where maximum pressures typically do not exceed $\simeq3$~GPa. Cubic-Zirconia was therefore chosen as the primary material for the developments presented in this paper.

\begin{figure}[htb]
\centering
\includegraphics[width=1.0\linewidth]{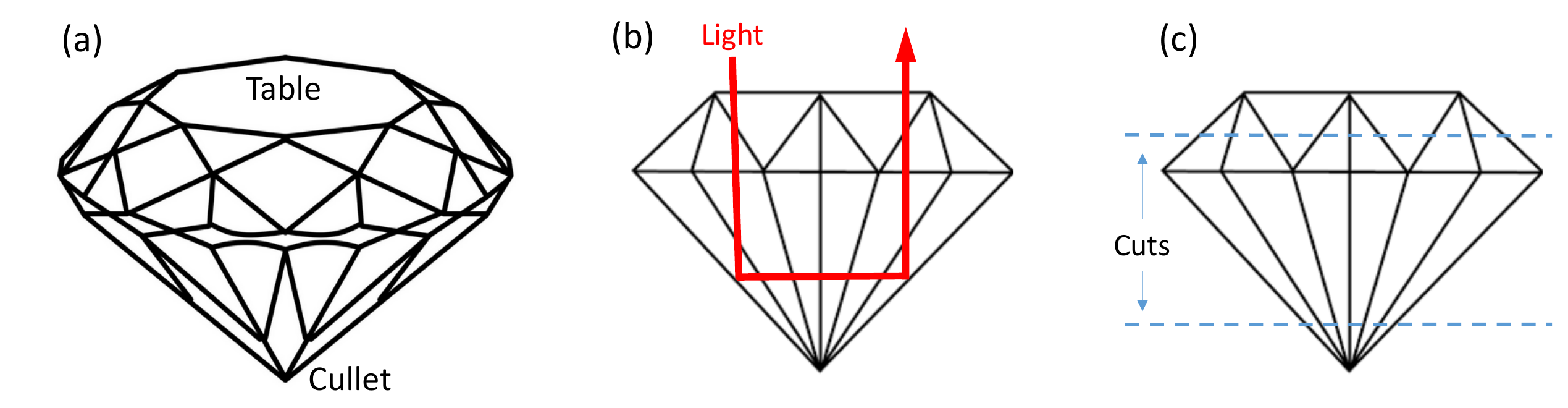}
\caption{(a) Schematic representation of a commercial gem stone in the `brilliant-cut' shape. (b) The  optical path inside the gem stone. In order to allow the shining light to pass through the crystal, the cullet must be flattened. (c) The preparation of the optical window, which requires the crystal truncation on both the table and cullet sides.}
\label{fig:Anvils preparation}
\end{figure}

Figure~\ref{fig:Anvils preparation}~(a) shows schematically the commercially available gem stone in the so-called `brilliant-cut' shape. The sharp and the flat parts of the crystal are called the `cullet' and the `table', respectively. The 'table' is utilised as a light entrance window.  One of the features of the `brilliant-cut' shape is that the incident light is fully reflected inside the crystal [see Fig.~\ref{fig:Anvils preparation}~(b)]. In our experiments, however, the light should be able to leave the crystal at the cullet side, so the `cullet' end of the crystal must be flattened [truncated, see Fig.~\ref{fig:Anvils preparation}~(c)].

On the table side, however, the crystals must be truncated for a slightly different motivation. The crystals, used in our optical setup, have a size of 4~mm. As delivered, the average table diameter is $\simeq 2.2$~mm, which is only marginally larger than the 1.5~mm hole in the piston. One clearly needs to increase the contact area between the flat part of the crystal and the piston surface [see Fig.~\ref{fig:Anvils preparation}~(c)].

\begin{figure}[htb]
\centering
\includegraphics[width=1.0\linewidth]{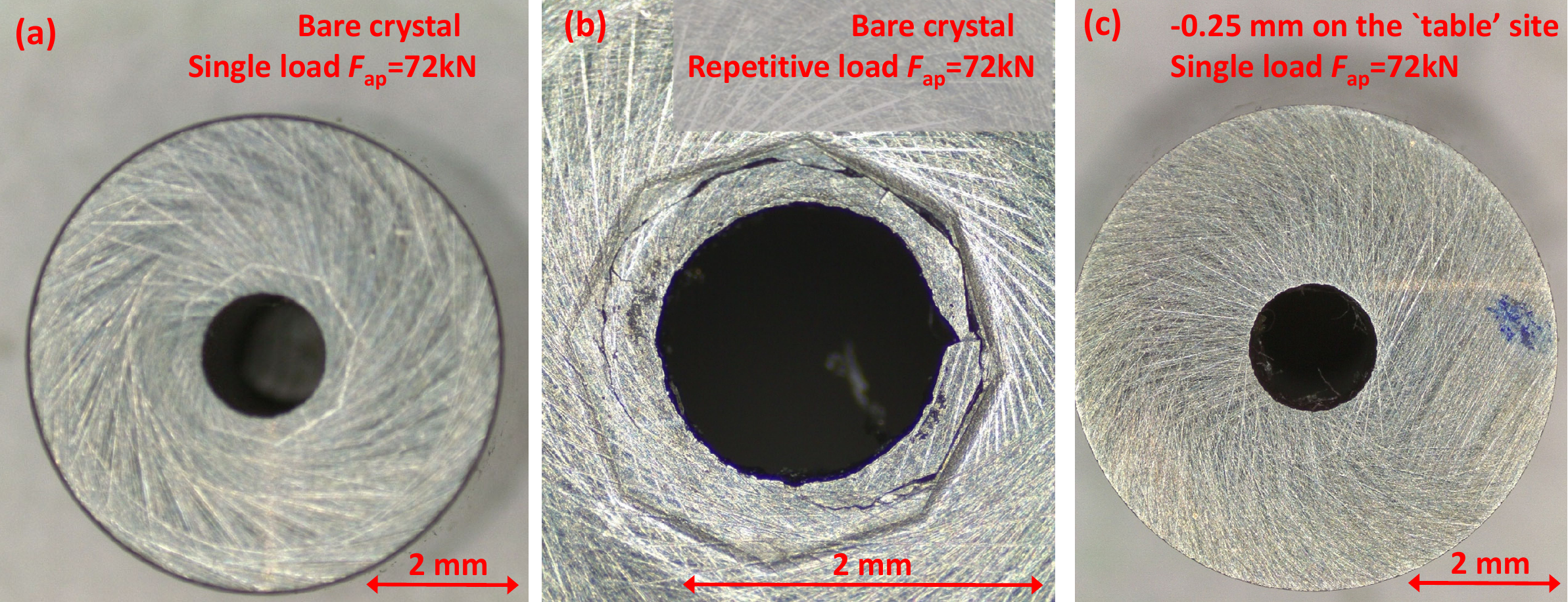}
\caption{(a) The top surface of WC piston after applying the force $F_{\rm ap}\simeq 72$~kN to the bare (not truncated) cubic-Zirconia crystal. The stamp around the center hole corresponds to the dimension of the crystal's `table'. (b) The top surface of WC piston after applying the repetitive load of $F_{\rm ap}\simeq 72$~kN to the bare cubic-Zirconia crystal. (c) The same as in panel (a) but for the crystal with 0.25~mm truncated `table' part. }
\label{fig:Pistons_below_crystals}
\end{figure}

Figure~\ref{fig:Pistons_below_crystals} shows the piston surface when the `table' part of the 4~mm crystal was not cut [panels (a) and (b)] and when it was cut by 0.25~mm [panel (c)]. Obviously, in the case of a bare crystal, even a single load with the maximum applied force $F_{\rm ap}\simeq 72$~kN (which corresponds to the applied pressure $p_{\rm ap}=F_{\rm ap}/S\simeq 2.5$~GPa, $S$ on the piston's top area) leads to the appearance of a clear step on the surface of the WC piston [Fig.~\ref{fig:Pistons_below_crystals}~(a)], while the repetitive load leads to damage to the piston around the central hole [Fig.~\ref{fig:Pistons_below_crystals}~(b)]. The use of the similar crystal, but truncated by $\simeq 0.25$~mm on the table side, increases the contact area between the piston and the crystal and leads to an almost complete disappearance of the crystal's step [Fig.~\ref{fig:Pistons_below_crystals}~(c)].

\subsection{Teflon mushroom}\label{sec:Teflon-Musroom}

The teflon mushroom with the conical entrance for the optical window [Fig.~\ref{fig:double-volume pressure cell}~(b)], plays a dual role: \\
(i) The top part of the mushroom's conical entrance works as a pressure volume for the optical probe (see Fig.~\ref{fig:concept_of_double-volume pressure cell}). \\
(ii) The teflon mushroom seals pressures in both the sample and optical volumes.

The pressure in the sample space is sealed as described in Ref.~\onlinecite{Shermadini_HPR_2017}. The initial pressure seal, up to $p\simeq 0.6-0.9$~GPa, is made by the rubber o-ring. With the pressure increase, the teflon mushroom deforms and it fills the volume between the beryllium copper (Be-Cu) sealing ring and the top area of the WC piston. The role of the BeCu ring is to prevent the flow of the teflon further down, {\it i.e.} to enter the area between the pressure cell channel and the walls of the WC piston.
The seal of the `optical volume' occurs simultaneously with the seal of the sample space. The teflon mushroom, being deformed by the applied pressure, seals also the optical volume, {\it i.e.} the small volume on top of the optical window, which is filled with the fluorescing material and the pressure transmitting media.

It is worthwhile to note that, after opening the pressure cell, both the piston and the optical window came out stuck together as one piece. There are two possible ways to proceed:\\
(i) The mushroom/window piece could be disintegrated (opened) and the assembly of a 'new` mushroom with an 'old` optical window might be used in the next experiment.   \\
(ii) The mushroom/window assembly remains unopened and is used in the next experiment. \\
The advantage of method (ii) is that there is no need to reload the 'optical` volume. Experiments reveal that the mushroom/window assembly might be reused approximately 3-4 times.

\section{${\rm SrB}_{4}{\rm O}_7:(0.01 {\rm Sm}^{2+},\; 0.03 {\rm Eu}^{2+}$) optical pressure sensor}\label{sec:SrB4O7}

The fluorescence method is widely used for in-situ pressure determination.\cite{Eremets_book_1996, Klotz_book_2013} For measurements in anvil cells, a small amount of the fluorescence-emitting material (ruby, SrB$_4$O$_7$, {\it etc.}) is, typically, placed inside the cell together with the sample.
In the clamp-type cells, one either enters with the fiber(s) inside the main sample volume,\cite{Koyama_RSI_2007} or places the optical calibrant between the piston and the teflon cup containing the sample.\cite{Podlesnyak_HPR_2018}

\begin{figure*}[htb]
\centering
\includegraphics[width=0.8\linewidth]{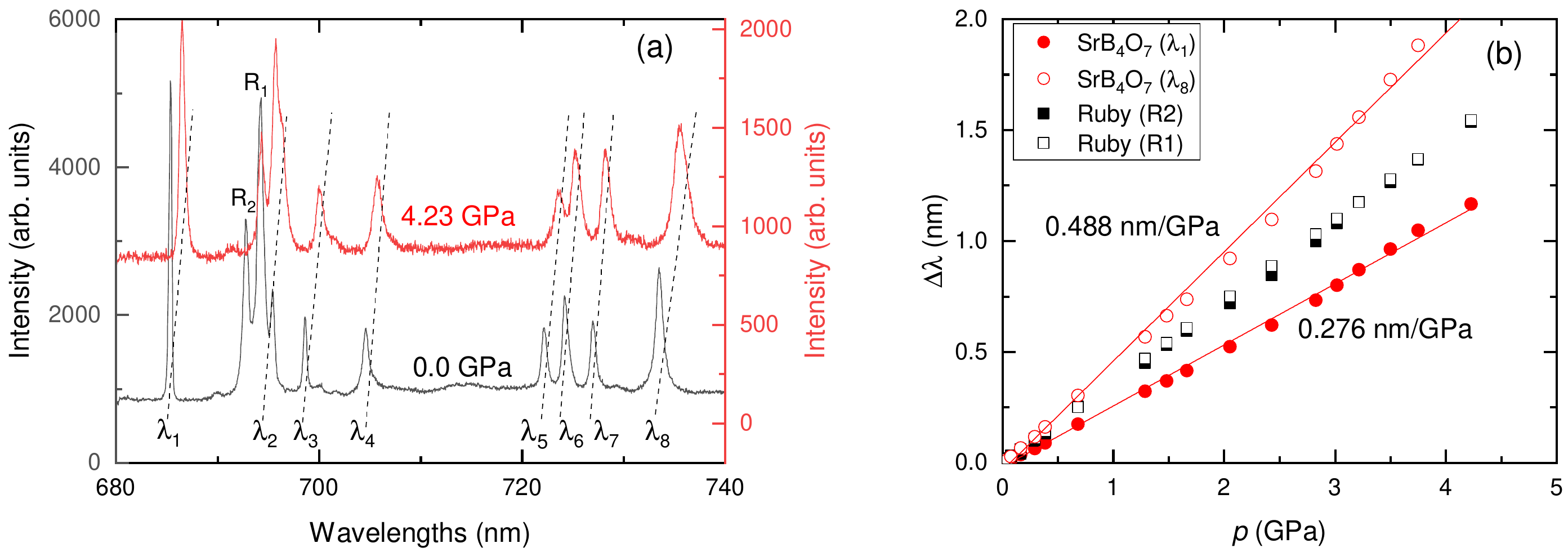}
\caption{(a) The fluorescence spectra of ruby and SrB$_{4}$O$_7$:(0.01 Sm$^{2+}$, 0.03 Eu$^{2+}$) collected at room temperature in diamond anvil cell at ambient pressure ($p=0.0$, black curve) and at $p=4.23$~GPa (red curve). The fluorescence spectra of ruby consists of two lines (R1 and R2).  The fluorescence spectra of SrB$_{4}$O$_7$:(0.01 Sm$^{2+}$, 0.03 Eu$^{2+}$) consists of eight lines (from $\lambda_1$ to $\lambda_8$). The dashed lines represent the pressure induced shifts of fluorescence lines.  (b) Pressure shifts of $\lambda_1$ and $\lambda_8$ lines of SrB$_{4}$O$_7$ and R1 and R2 lines of ruby. The solid lines are linear fits to $\lambda_1(p)$ and $\lambda_8(p)$ data (see text for details). }
\label{fig:SrB4O7 spectra}
\end{figure*}

In the present study, we followed the way outlined by Podlesnyak {\it et. al.}\cite{Podlesnyak_HPR_2018} who suggested to use the samarium-doped strontium tetraborate (SrB$_4$O$_7$:Sm$^{2+}$) as a fluorescence sensor for clamp-type pressure cells. In accordance with the literature,\cite{Podlesnyak_HPR_2018, Datchi_JApPh_1997, Lacam_JApPh_1989} SrB$_{4}$O$_7$:Sm$^{2+}$ has several advantages compared to ruby (the most commonly used optical sensor). In particular, the main excited line, $^{7}D_0- \; ^5F_0$, has a weak temperature dependence $d\lambda/dT \simeq -10^{-4}$~nm/K compared to
$d\lambda/dT \simeq 6.8\cdot 10^{-3}$~nm/K for the ruby excitation.\cite{Podlesnyak_HPR_2018, Datchi_JApPh_1997, Lacam_JApPh_1989, Barnett_RSI_1973} Moreover, the position of the narrow and well isolated $^{7}D_0- \; ^5F_0$ Sm$^{2+}$  line  could be determined more accurately than the strongly overlapping R1 and R2 peaks in the case of ruby. In addition, in Ref.~\onlinecite{Zheng_JMCh_2020} it was demonstrated that incorporating Eu$^{2+}$ into the crystal structure of SrB$_{4}$O$_7$:Sm$^{2+}$ may lead to an enhancement ($\simeq 60$ times) of the Sm$^{2+}$ emission.

 \begin{table}[htb]
     \centering
     \caption{Wavelengths of eight fluorescence lines of SrB$_{4}$O$_7$:(0.01 Sm$^{2+}$, 0.03 Eu$^{2+}$) at ambient pressure [$\lambda_i(p=0)$, $i=1, ..8$] and the corresponding pressure induced shifts [$d\lambda_i/dp$]. The errors are obtained from the linear fits to $\lambda_{i}(p)$ data [see Fig.~\ref{fig:SrB4O7 spectra}~(b)]. The fourth and the fifth columns show the similar data for SrB$_{4}$O$_7$:(0.05 Sm$^{2+}$) from Ref.~\onlinecite{Lacam_JApPh_1989}. \\}
     \begin{tabular}{c|cc|ccc}
 Line &\multicolumn{2}{c|}{SrB$_{4}$O$_7$:(0.01 Sm$^{2+}$, 0.03  Eu$^{2+}$)}&\multicolumn{2}{c}{SrB$_{4}$O$_7$:0.05 Sm$^{2+}$, [\onlinecite{Lacam_JApPh_1989}]}\\
 \cline{2-5}
 &  $\lambda_i(p=0)$ & $d\lambda_i/dp$ &$\lambda_i(p=0)$ &$d\lambda_i/dp$   \\
&nm& nm/GPa&nm& nm/GPa\\
 \hline
 $\lambda_1$  & 685.359(6) &0.276(3)&685.41&0.255 \\
 $\lambda_2$  & 695.416(9) &0.244(5)&695.53&0.230 \\
 $\lambda_3$  & 698.604(9) &0.360(5)&698.67&0.295 \\
 $\lambda_4$  &704.62(1)   &0.279(4)&704.66&0.250 \\
 $\lambda_5$  &722.151(5)  &0.336(3)&722.24&0.325 \\
 $\lambda_6$  &724.234(4)  &0.251(3)&724.29&0.215\\
 $\lambda_7$  &726.971(9)  &0.292(6)&727.05&0.255 \\
 $\lambda_8$  &733.47(1)   &0.488(7)&733.64&0.450 \\
      \end{tabular}
      \label{tab:The frequency table}
 \end{table}

SrB$_{4}$O$_7$:(0.01 Sm$^{2+}$, 0.03 Eu$^{2+}$) was prepared in a way described in Ref.~\onlinecite{Zheng_JMCh_2020}. The calibration of SrB$_{4}$O$_7$:(0.01 Sm$^{2+}$, 0.03 Eu$^{2+}$)  was performed in a diamond anvil cell by covering a pressure range from 0 to $\simeq4.2$~GPa. In this experiment a small ruby chip was placed inside the cell together with SrB$_{4}$O$_7$. The pressure was determined by measuring the pressure induced shift of the R1 ruby line: $ d \lambda_{\rm R1}/ d p =0.365(5)$~nm/GPa.\cite{Datchi_JApPh_1997, Piermarini_JG_1975}

Figure~\ref{fig:SrB4O7 spectra}~(a) shows two fluorescence spectra collected in a diamond anvil cell at ambient pressure ($p=0.0$, black line) and at $p=4.23$~GPa (red line). Two ruby and eight SrB$_{4}$O$_7$ fluorescence lines are denoted by `R' and `$\lambda$' letters, respectively. With increasing pressure, all lines shift to the higher wavelengths. The comparison of the pressure induced shift of $\lambda_1$ and $\lambda_8$ fluorescence lines of SrB$_{4}$O$_7$ with the $R1$ and $R2$ lines of ruby is presented in Fig.~\ref{fig:SrB4O7 spectra}~(b). It is interesting to note that the pressure shift of $\lambda_8$ line [$d\lambda_8/dp= 0.488(7)$~nm/GPa] exceeds the corresponding pressure coefficient of the R1 line of ruby [$ d \lambda_{\rm R1}/ d p =0.365(5)$~nm/GPa] by more than 30\%.

The results of linear fits to the pressure dependencies of the SrB$_{4}$O$_7$:(0.01 Sm$^{2+}$, 0.03 Eu$^{2+}$) fluorescence lines are summarised in Table~\ref{tab:The frequency table}. For comparison, the data for SrB$_{4}$O$_7$:(0.05 Sm$^{2+}$) from Ref.~\onlinecite{Lacam_JApPh_1989} are also included. Obviously, the line positions at ambient pressure and the linear coefficients determined in this work are slightly different from those obtained by Lacam {\it et al.} in Ref.~\onlinecite{Lacam_JApPh_1989}. Note also, that $d\lambda_1/dp=0.241$~nm/GPa was reported by Podlesnyak {\it et al.} in Ref.~\onlinecite{Podlesnyak_HPR_2018}. The clear difference between our data and that mentioned in literature imply that the SrB$_{4}$O$_7$-based optical probes are not universal. One needs to therefore perform individual calibrations for each newly synthesized SrB$_{4}$O$_7$ batch.
\begin{figure}[htb]
\centering
\includegraphics[width=1.0\linewidth]{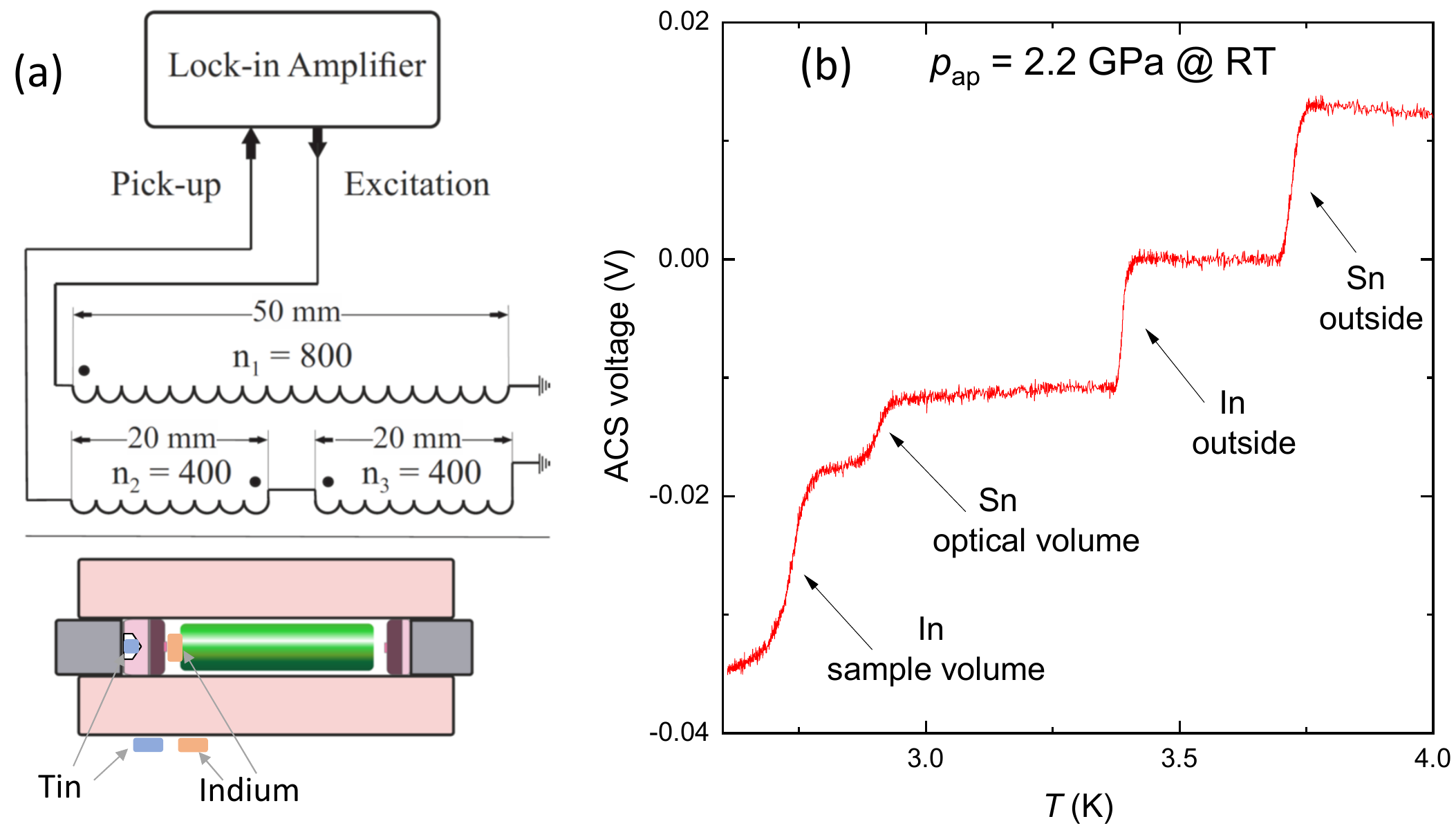}
\caption{(a) The ac susceptibility setup used to simultaneously pressure measurements inside the 'sample` and the 'optical` volumes. The ACS system with the  lock-in amplifier and coils (one excitation and two pick-up) is the same as described in Ref.~\onlinecite{Khasanov_HPR_2016}. The pressure probes (pieces of Sn and In) are placed inside the  'sample` and the 'optical` volumes, as well as attached to the pressure cell body. (b) The ACS curve measured after closing the cell at $p_{\rm ap}\simeq 2.2$~GPa. Arrows mark the superconducting transitions of 4 pressure probes.}
\label{fig:Indium-Sn}
\end{figure}

\begin{figure*}[htb]
\centering
\includegraphics[width=0.7\linewidth]{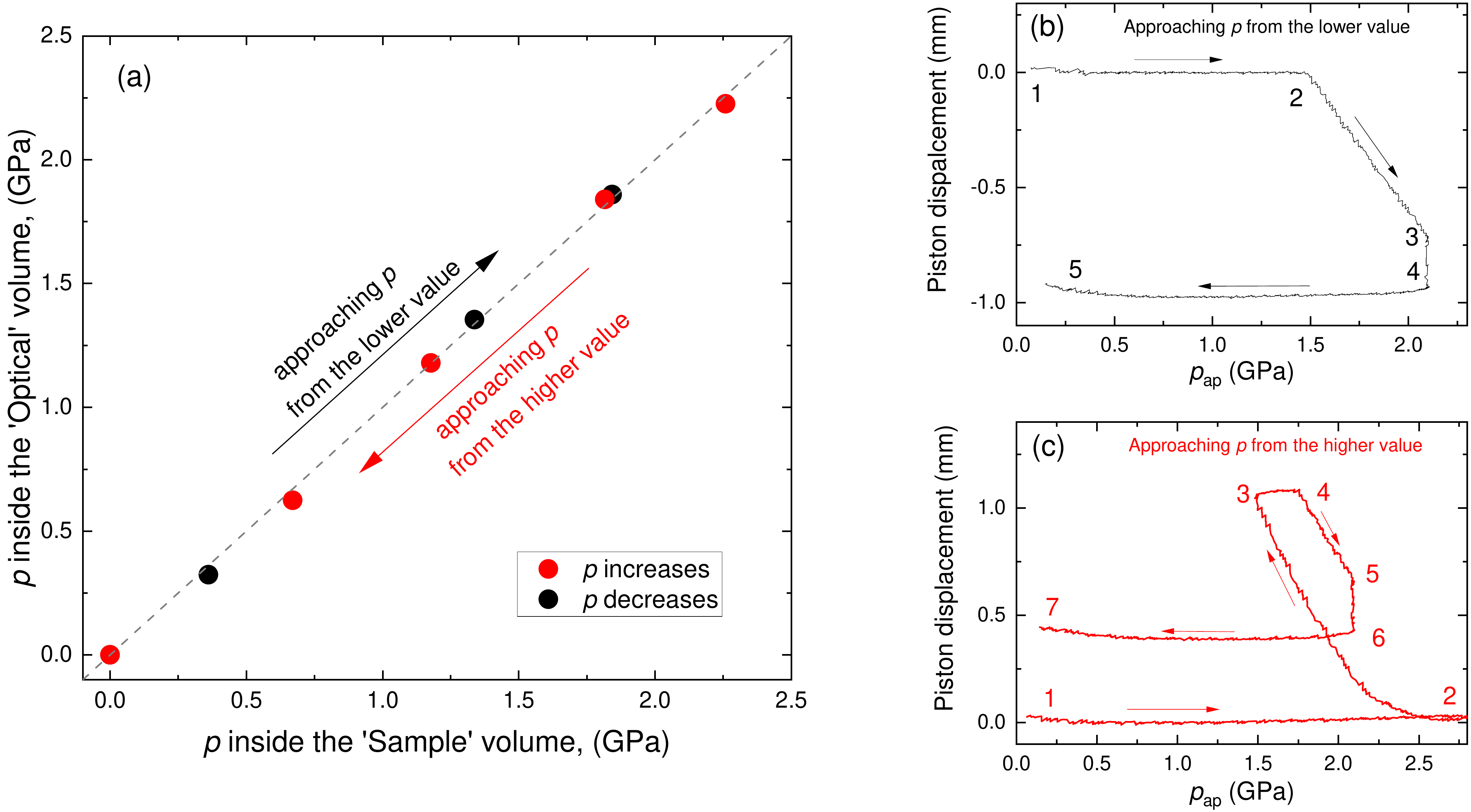}
\caption{ (a) Dependence of the pressure inside the 'optical` volume on the pressure inside the 'sample` volume. The dashed line represents the one-to-one correspondence between both pressures. (b) The pressure cell loading curve representing the process of approaching the final pressure from the lower value. The pressure cell was initially closed at $\simeq 1.5$~GPa, so the piston is not displaced until $p_{\rm ap}$ reaches a value of $1.5$~GPa (the path from Point 1 to 2). The piston moves inside the cell with the $p_{\rm ap}$ increase from $\simeq 1.5$ to $2.1$~GPa, where the cell is closed by tightening the fixing bolt (the path 2--3--4). The force applied to the bolt leads to an additional movement of the piston by $\simeq0.2$~mm further down (the path 3--4). The position of the piston does not changed by releasing $p_{\rm ap}$ down to 0 (path 4--5). (c) The pressure cell loading curve representing the process of approaching the final pressure from the higher value. The fixing bolt is unscrewed by reaching point 2. By decreasing $p_{\rm ap}$, the piston moves outside of the pressure cell channel (the path 2--3). The rest of the process (the paths 3--4, 4--5, 5--6, and 6--7) is the same as described in panel (b).   }
\label{fig:pressures_Sample-Optical}
\end{figure*}

\section{Test experiments}\label{sec:experiments}

\subsection{Pressures inside the `sample' and the `optical' volumes}

In order to use the double-volume clamp cell in real experiments, one needs to ensure that pressures inside the `sample' and the `optical' volumes are equal. The importance of such confirmation becomes obvious by taking into account the details of the pressure application process. The force from the external press is transferred through the top piston and the top seal system to the sample chamber and then, via the teflon mushroom, into the optical volume [see Fig.~\ref{fig:double-volume pressure cell}~(a)]. The pressure seal
can lead to a large amount of friction due to the piston sliding past the cell,\cite{Walker_RSI_1999} so the real pressure inside the cell becomes smaller compared to the applied one ($p<p_{\rm ap}$) during the cell pressurisation process, and it is higher than $p_{\rm ap}$ ($p>p_{\rm ap}$) during the cell depressurisation procedure.\cite{Khasanov_ThreeWall_Arxiv_2021}

The setup of our test experiments is shown schematically in Fig.~\ref{fig:Indium-Sn}~(a). A standard $\mu$SR double-wall piston-cylinder pressure cell [similar to the one shown in Fig.~\ref{fig:double-volume pressure cell}~(a) and described in Refs.~\onlinecite{Khasanov_HPR_2016, Shermadini_HPR_2017}] was used. Two pressure indicators (small pieces of elemental metals, Sn and In) were placed inside of the `optical' and the `sample' volumes. Another two `reference' probes were attached to the pressure cell body. The piston with the optical entrance (1.5~mm central hole) was replaced to the bare one and the optical window was not used.
The pressure was determined by measuring shifts of the superconducting transition temperatures ($T_{\rm c}$'s) of In and Sn from their ambient pressure values [$T_{\rm c}^{\rm In}(p=0)\simeq3.405$~K and $T_{\rm c}^{\rm Sn}(p=0)\simeq3.717$~K] by means of the ac susceptibility (ACS) technique. The design and operation of the ACS measuring setup used in our study is described in detail in Ref.~\onlinecite{Shermadini_HPR_2017}.

Figure~\ref{fig:Indium-Sn}~(b) gives an example of ACS measurements at $p_{\rm ap}\simeq 2.2$~GPa applied at a room temperature. The superconducting transitions of all 4 pressure indicators are clearly visible and are well separated from each other. For pressure determination the following equations were used:\cite{Eiling_IJMP_1981}
\begin{equation}
T_{\rm c}^{\rm In}(p)= T_{\rm c}^{\rm In}(0)-0.3812\; p+0.0122\; p^2
\label{eq:In Tc}
\end{equation}
and
\begin{equation}
T_{\rm c}^{\rm Sn}(p)= T_{\rm c}^{\rm Sn}(0)-0.4823\; p+0.0207\;p^2.
\label{eq:Sn Tc}
\end{equation}
Here, $T_{\rm c}(0)$ is the superconducting transition temperature at $p=0$. In Fig.~\ref{fig:Indium-Sn}~(b), {\it e.g.}, $T_{\rm c}^{\rm In}(0)$ and  $T_{\rm c}^{\rm Sn}(0)$ might be obtained from the transitions marked as `In outside' and `Sn outside', respectively.  $T_{\rm c}^{\rm In}(p)$ and $T_{\rm c}^{\rm Sn}(p)$ correspond to the superconducting transitions of In and Sn pressure probes mounted into the `sample' and the `optical' volumes, respectively.

\begin{figure*}[htb]
\centering
\includegraphics[width=0.8\linewidth]{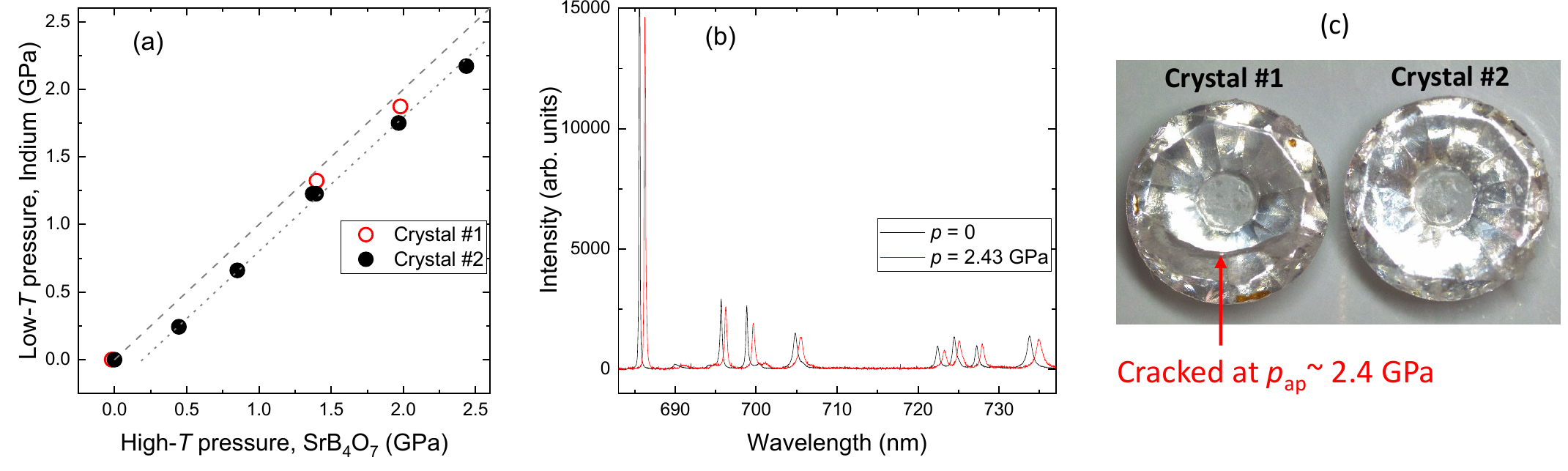}
\caption{(a) Dependence of the low-temperature pressure value (at $T\simeq 3$~K, as is determined from the $T_{\rm c}$ shift of indium) on the high-$T$ pressure (as is obtained in florescence measurements at room temperature). The pressure difference between the high-$T$ and the low-$T$ pressure values is $\simeq 0.2$~GPa, which is typical for piston-cylinder cells and it is caused by the thermal contraction of the pressure transmitting media (Daphne 7373 oil in our case). (b) The fluorescence spectra of SrB$_{4}$O$_7$:(0.01 Sm$^{2+}$, 0.03 Eu$^{2+}$) ap $p=0$ (black curve) and $p=2.43$~GPa (red curve). (c) Cubic-Zirconia crystals used in this experiment. The Crystal \#1 was cracked at $p_{\rm ap}\simeq 2.4$~GPa, while the Crystal \#2 survived after several loading cycles with the applied pressure reaching $\simeq 2.8$~GPa}
\label{fig:Indium-SrB4O7}
\end{figure*}

The dependence of the pressure inside the 'optical` volume as a function of the pressure inside the 'sample` volume is presented in Fig.~\ref{fig:pressures_Sample-Optical}~(a). The resulting pressures were approached in two different ways: from the lower (black points) and from the higher (red points) pressure values. The corresponding pressure cell loading curves representing the displacement of the piston as a function of the applied pressure are shown in panels (b) and (c).

The results presented in Fig.~\ref{fig:pressures_Sample-Optical} indicate how the double-volume concept as discussed in Sec.~\ref{sec:Motivation} works remarkably well. Indeed, pressures inside the 'Sample` and the 'Optical` volumes are the same within the experimental accuracy. There is no hysteresis if the final pressure reached from the previously set higher or lower values.

\subsection{Experiments with the fluorescence probe}

In order to test the optical measurement setup, as is presented in Fig.~\ref{fig:concept_of_double-volume pressure cell}, the In pressure indicator was placed inside the 'sample` chamber and a small drop of a mixture of SrB$_{4}$O$_7$:(0.01 Sm$^{2+}$, 0.03 Eu$^{2+}$) with the Daphne 7373 oil ($\simeq 0.5-1.0$~mm$^3$) was introduced into the 'optical` volume.
The results of these test experiments are shown in Fig.~\ref{fig:Indium-SrB4O7}. Pressures inside the `sample' and the 'optical` volumes were determined from the $T_{\rm c}$ shift of indium and the fluorescence lines of SrB$_{4}$O$_7$, respectively.
The fluorescence lines of SrB$_{4}$O$_7$:(0.01 Sm$^{2+}$, 0.03 Eu$^{2+}$) at ambient pressure and at $p\simeq 2.43$~GPa are shown in Fig.~\ref{fig:Indium-SrB4O7}~(b). The pressure induced shift of all eight fluorescence lines is clearly visible.

Figure~\ref{fig:Indium-SrB4O7}~(a) shows  that the pressure values determined by the use of indium as a pressure indicator are systematically lower ($\simeq 0.2$~GPa) than those obtained by means of SrB$_{4}$O$_7$ fluorescence.
Note that the pressure determined via the superconducting transition temperature of In corresponds to the low-temperature value of the pressure [$T\simeq 3$~K, see Fig.~\ref{fig:Indium-Sn}~(b)], while the fluorescence spectra of SrB$_{4}$O$_7$ were measured at a room temperature. The pressure drop by cooling is typical for piston-cylinder cells and it is caused, mainly, by the thermal contraction of the pressure transmitting media.\cite{Torikachvili_RSI_2015} In our case, 7373 Daphne oil was used, which is characterized by the pressure drop via cooling of the order of 0.15-0.2~GPa.\cite{Yokogawa_JAP_2007, Stasko_HPR_2020}

The closed and open symbols in Fig.~\ref{fig:Indium-SrB4O7}~(a) refer to the data taken by using two different cubic-Zirconia crystals (\#1 and \#2). Crystal \#1 was cracked at $p_{\rm ap}\simeq 2.4$~GPa, while Crystal \#2 survived after several loading cycles with the applied pressure reaching $p_{\rm ap}\simeq 2.8$~GPa. This implies that:\\
(i) the maximum pressure of commercially available cubic-Zirconia gemstones may vary from crystal to crystal and \\
(ii) at present conditions, the cubic-Zirconia crystals stay near their mechanical limit.

\section{Finite element analysis}\label{sec:FEA}

\begin{figure}[htb]
\centering
\includegraphics[width=1.0\linewidth]{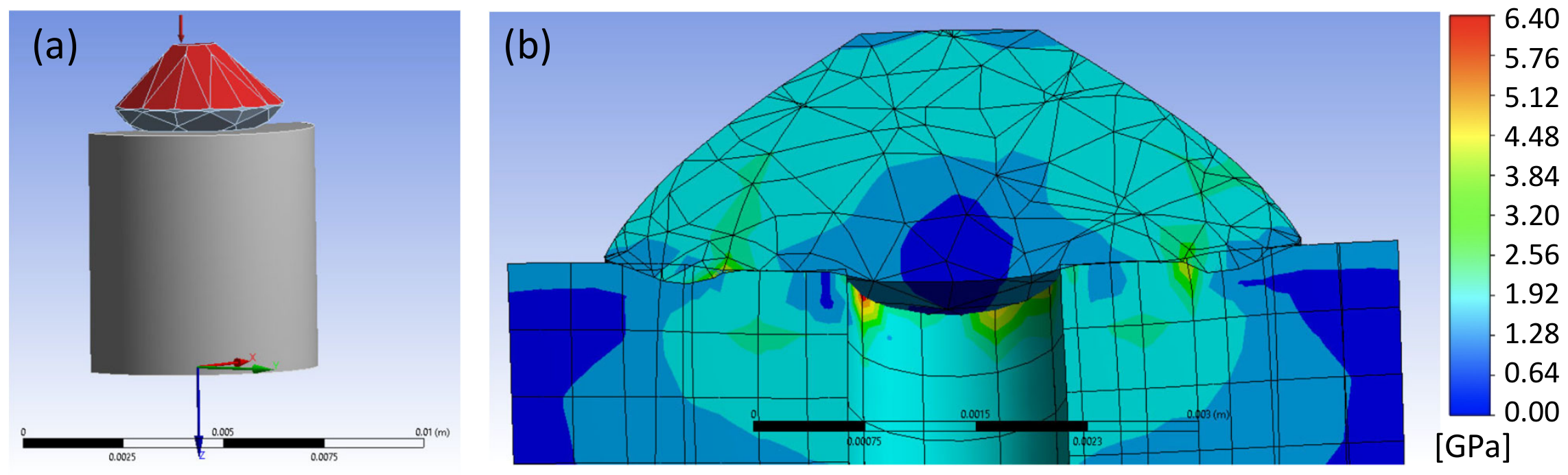}
\caption{(a) Finite element analysis model of the optical window and the WC piston assembly. A pressure of 1~GPa is applied to the top part of the optical window (the red colored area). (b) The distribution of mechanical stress in the contact area between the top of the WC piston and the optical window.}
\label{fig:FEA-model}
\end{figure}

To estimate the amount of mechanical stress on the optical window and the WC piston with 1.5~mm inner hole, the piston/window assembly was modeled using ANSYS finite element analysis software.\cite{ANSYS} The model was set up as shown in Fig.~\ref{fig:FEA-model}~(a). The external 1~GPa pressure was applied to the top part of the cubic-Zirconia crystal (the red colored area in  Fig.~\ref{fig:FEA-model}~(a)].

The results of the analysis are presented in Fig.~\ref{fig:FEA-model}~(b). The maximum load was found in the area around 1.5~mm hole, {\it i.e.}, exactly at the place where damages on the WC piston [Fig.~\ref{fig:Pistons_below_crystals}~(b)] and the crack on the Crystal \#1 [Fig.~\ref{fig:Indium-SrB4O7}~(c)] appeared. Although the maximum pressure in this area ($\sim6$~GPa) is far from the cubic-Zirconia pressure limit ($\simeq 16.7$~GPa, see Table~\ref{tab:The frequency table}), the failure of some optical windows may relate to the low quality of commercially available crystals. At the same time, the pressure around the optical entrance  exceeds the ultimate tensile stress of the tungsten carbide ($\simeq370$~MPa, Ref.~\onlinecite{Tungsten_carbide}) and results in an anvil-shaped indentation of WC piston [see also photos in Figs.\ref{fig:Pistons_below_crystals}~(a) and (b)].

The results of ANSYS simulations supports our experimental findings. They suggest that in order to reach repeatedly pressures above 2~GPa by using the optical setup described above, either the diameter of the hole inside the WC piston must be reduced from 1.5 down to {\it e.g.} 1~mm, or the optical window material must be replaced by 'stronger` Sapphire, Moissanite or Diamond crystals (see Table \ref{tab:Anvils}).

\section{Conclusions and outlook}\label{sec:conclusions}

To summarize, a double-volume clamp cell measurements setup was successfully designed and tested. With the present construction, the pressure inside the 'optical` and the 'sample` volumes were proved to be the same within the experimental accuracy.
The present concept could be used with almost all types of piston-cylinder cells and requires just a minor modification of the pressure sealing part.

The advantages of the presently proposed double-volume setup compared to the usual single volume approach are the following: \\
 1. The pressure calibrant stays in a separate volume and it is not mixed with the sample. This allows one to avoid any possible reaction between the sample and the calibrant material.\\
 2. The optical probe stays a few millimeters below the sample so that is not hit by neutrons or muons with properly collimated beams. \\
 3. The volume occupied by the optical probe is of the order of 0.2-0.5~mm$^3$ which is two orders of magnitude smaller than the volume available for the sample in standard muon and neutron pressure cells.

The use of cubic-Zirconia windows, produced from commercially available gemstones, has been shown to be a reliable and simple approach. On the other hand, the mechanical failure of some cubic-Zirconia crystals at pressures exceeding 2~GPa indicates the diverse quality of commercially available gemstones. Sapphire, Moissanite or Diamond crystals with higher mechanical strength could be used as a more reliable and expensive alternative. An alternative solution could the reduction of the optical entrance into the tungsten carbide piston from currently used 1.5~mm, down to 1.0~mm or even smaller. It might be also helpful to increase the contact area between the piston and the optical window by using truncated ball-shaped crystals instead of crystals shaped in a 'brilliant-cut`.

\begin{acknowledgments}
The work was performed at the Swiss Muon Source (S$\mu$S), Paul Scherrer Institute (PSI, Switzerland). The work of R.G. was supported by the Swiss National Science Foundation (SNF Grant No. 200021-175935). R.K. acknowledges the help of Charles Hillis Mielke III for careful reviewing the manuscript.
 \label{sec:Acknowledgement}
\end{acknowledgments}

\end{document}